\documentclass[runningheads]{llncs}
\usepackage[T1]{fontenc}
\usepackage{graphicx}
\usepackage{booktabs}
\usepackage{multirow}
\usepackage[normalem]{ulem}
\usepackage[sectionbib, numbers]{natbib}
\useunder{\uline}{\ul}{}

\usepackage{amsmath}
\usepackage{framed}
\usepackage{mdframed}
\usepackage{lipsum}

\usepackage[colorlinks=true, linkcolor=blue, urlcolor=blue]{hyperref}

\usepackage{footnote}
\usepackage{enumitem}
\usepackage{longtable}
\usepackage{color}
\usepackage[table,dvipsnames]{xcolor}
\usepackage{adjustbox}
\usepackage{diagbox}
\usepackage{array}
\newcolumntype{P}[1]{>{\centering\arraybackslash}p{#1}}
\usepackage{makecell}

\usepackage{xcolor}

\usepackage{amsfonts}

\begin{document}
\title{Automatic Large Language Models Creation of Interactive Learning Lessons}

\titlerunning{Automatic Large Language Models Creation of Interactive Learning Lessons}

\author{Jionghao Lin\inst{1,2,3} \and
Jiarui Rao\inst{2} \and
Sandy Yiyang Zhao\inst{2} \and
Yuting Wang\inst{2} \and
Ashish Gurung\inst{2} \and 
Amanda Barany\inst{4} \and 
Jaclyn Ocumpaugh\inst{4} \and 
Ryan S. Baker\inst{4} \and 
Kenneth R. Koedinger\inst{2}}
\authorrunning{J. Lin et al.}

\institute{The University of Hong Kong, Pokfulam Rd, Hong Kong, China\\
\email{jionghao@hku.hk} \and
Carnegie Mellon University,	Pittsburgh, PA 15213, USA
\and
Monash University, Wellington Rd, Clayton VIC 3800, Australia
\and
University of Pennsylvania, Philadelphia, PA 19104, USA
}

\begin{frontmatter}
\maketitle
%

\begin{abstract}


We explore the automatic generation of interactive, scenario-based lessons designed to train novice human tutors who teach middle school mathematics online. Employing prompt engineering through a Retrieval-Augmented Generation approach with GPT-4o, we developed a system capable of creating structured tutor training lessons. Our study generated lessons in English for three key topics\textemdash Encouraging Students' Independence, Encouraging Help-Seeking Behavior, and Turning on Cameras\textemdash using a task decomposition prompting strategy that breaks lesson generation into sub-tasks. The generated lessons were evaluated by two human evaluators, who provided both quantitative and qualitative evaluations using a comprehensive rubric informed by lesson design research. Results demonstrate that the task decomposition strategy led to higher-rated lessons compared to single-step generation. Human evaluators identified several strengths in the LLM-generated lessons, including well-structured content and time-saving potential, while also noting limitations such as generic feedback and a lack of clarity in some instructional sections. These findings underscore the potential of hybrid human–AI approaches for generating effective lessons in tutor training.

\end{abstract}
\end{frontmatter}

\section{Introduction}

One-on-one human tutoring is widely recognized as an effective strategy for enhancing student learning, with substantial evidence highlighting its positive impact on learning outcomes \cite{ nickow2020impressive}. However, scaling human tutoring remains a significant challenge due to the scarcity of skilled tutors \cite{nickow2020impressive}. To address this issue, professional development programs have been developed to equip novice and nonprofessional tutors with the skills needed for effective instruction \cite{nickow2020impressive}. Online scenario-based training has emerged as a promising approach, simulating real-life tutoring situations to prepare novice tutors for their roles \cite{thomas2023tutor}. Central to scenario-based tutor training program is well-constructed lessons that offer specific instructional strategies and practical guidance for adapting human tutoring to diverse learner needs. 
However, developing such lessons manually requires significant time and expertise, making the lesson design process costly and difficult to scale. These challenges underscore the urgent need for artificial intelligence (AI) technologies for automatic lesson generation, which can streamline lesson development and expand access to effective tutor training programs.


Lesson generation using AI techniques has been investigated in previous research to address the challenges of the time-consuming and labor-intensive process of creating educational materials. Early approaches employed rule-based systems to generate quiz questions, providing a foundational method but lacking adaptability and depth \cite{wiemer1998foundations, koedinger2006cognitive}. Subsequent approaches introduced template-based methods and early AI models, which improved the variety and relevance of generated content \cite{paladines2020systematic}. However, these approaches still faced significant limitations, including inflexibility and a lack of contextual understanding \cite{paladines2020systematic}.

Recently, large language models (LLMs) demonstrate exceptional capabilities in generating human-like text, leveraging pretraining on extensive datasets \cite{radford2018improving}. These capabilities enable LLMs to potentially address the shortcomings of earlier approaches by producing nuanced, contextually relevant, and adaptable content, making them particularly well-suited for educational purposes. 
However, generating comprehensive lessons for tutor training remains a complex task. A lesson typically consists of multiple interconnected segments, including well-defined learning objectives and quiz. Each segment serves a specific instructional purpose and must align with the overall pedagogical goals of the lesson. 
Given the complexity of lesson design, generating all components in a single step risks overwhelming the model, resulting in shallow or disjointed outputs \cite{khot2022decomposed}. 

To address this, we adopt a prompting strategy similar to \cite{wu2022ai}, chaining multiple LLM steps together\textemdash where the output of one step serves as the input for the next\textemdash and has been shown to enhance the quality of task outcomes. Research highlights the benefits of such decomposition strategies for tasks requiring multi-step reasoning, as they help improve the quality of generated content \cite{wu2022ai, khot2022decomposed}. However, it remains unknown how extensively a task should be decomposed to ensure optimal performance, since overly complex sub-goals might still overwhelm the LLM \cite{li2024towards}. In the context of lesson generation, this challenge becomes particularly relevant, as lessons consist of multiple interdependent components that must align both structurally and pedagogically. It remains unclear how the number of segments or the level of decomposition impacts the quality, coherence, and pedagogical rigor of the generated lessons.





The demand for high-quality lesson content in tutor training programs remains substantial; however, the development of content specifically tailored for tutor training has been relatively under-explored. Our study developed a lesson content generation system that leverages prompt engineering with LLMs. The system was designed to generate scenario-based lessons for training novice tutors who teach middle school mathematics in online environments. To evaluate the effectiveness of the generated lessons, we collaborated with human instructional designers, who evaluated the lesson quality using a comprehensive rubric. This evaluation aimed to determine how effectively LLMs can generate lesson content that meets the pedagogical and practical needs of tutor training. Based on this investigation, our study proposed two \textbf{R}esearch \textbf{Q}uestions (\textbf{RQs}): \textbf{RQ1:} \textit{How does the level of task decomposition influence the quality of tutor training lessons generated by LLMs?} \textbf{RQ2:} \textit{How does the quality of lessons generated by LLMs compare to those created by expert human lesson designers?}

\section{Related Work}
\vspace{-2mm}
\subsection{Scenario-based Lesson Development for Tutor Training}
\vspace{-1mm}

Scenario-based lessons are a key component of effective tutor training, providing a realistic context for tutors to apply their skills \cite{chhabra2022evaluation}. These lessons simulate authentic tutoring interactions, helping tutors understand and navigate common challenges they might face. 
By working through scenarios that mimic real-world situations, tutors develop the flexibility needed to effectively support diverse learners. Additionally, lessons focusing on advocacy, empathy, and cultural competence are important for preparing tutors to work with diverse student populations \cite{thomas2023tutor}. These competencies enable tutors to deliver academic content and address students' emotional well-being and equity-related challenges, fostering supportive learning environments \cite{chhabra2022evaluation}.

Lessons structured around ``learning by doing'' methodologies, such as the modified Predict-Observe-Explain (POE) model, have shown significant success in enhancing tutor preparedness \cite{koedinger2015learning}. In the scenario-based lessons, tutors first engage with a scenario by predicting the best response and then explaining their rationale. This approach fosters active engagement and critical reflection, helping tutors develop a deeper understanding of effective tutoring strategies \cite{aleven2002effective}. 
Moreover, incorporating a combination of multiple-choice questions (MCQs) and open-ended questions within these lessons has proven effective for developing both basic and advanced tutoring skills \cite{butler2018multiple}. MCQs help in assessing foundational knowledge in an efficient manner, while open-ended questions encourage deeper cognitive engagement, enabling tutors to articulate their understanding \cite{gurung2024multiple}. 

\vspace{-3mm}

\subsection{Lesson Generation}
\vspace{-1mm}

Lessons designed for tutor training are essential for equipping tutors with the skills needed to support diverse learners. However, manually crafting these lessons can be both time-consuming and labor-intensive, particularly when scaling tutor training programs to meet growing demand \cite{nickow2020impressive, thomas2023tutor}. Automated lesson generation approaches have emerged as a potential solution to these challenges, offering benefits such as time savings, scalability, and consistency in instructional content. Traditional methods of automated learning content generation, such as rule-based approaches or template-based approaches \cite{diwan2023ai}, often lack the flexibility and contextual depth. These limitations highlight the need for more advanced approaches to generating structured, pedagogically rich lessons.

The rise of LLMs has brought significant advancements to content generation in education. LLMs have demonstrated strong performance across a range of educational tasks, including generating feedback \cite{10.1007/978-3-031-42682-7_19, dai2024assessing}, course learning topics \cite{moein2024beyond} and questions \cite{nguyen2022towards}. For instance, GPT models have been used to provide feedback on open-ended student responses, producing actionable suggestions that support learning \cite{dai2024assessing}. 
Despite the growing interest in LLMs for education, their use in generating structured, scenario-based lessons tailored to tutor training remains under-explored. Given the unique requirements of tutor training, the potential for LLMs to automate high-quality lesson generation at scale presents a significant opportunity.



\section{Methods}
\vspace{-2mm}

\subsection{Lesson Structure from a Tutor Training Platform}
\label{lesson_structure}
\vspace{-1mm}
To develop training lessons for novice human tutors teaching middle school mathematics online, our study builds upon an existing tutor training program structured around the SMART framework demonstrated in \cite{chhabra2022evaluation, thomas2023tutor}, which encompasses five key competencies critical for successful tutoring: \textbf{S}ocial-emotional Learning, \textbf{M}astery of Content, \textbf{A}dvocacy, \textbf{R}elationship-building, and \textbf{T}echnology. All lessons in our study were generated in English.
The lessons are divided into five main sections, each designed to support specific aspects of tutor training:
    \vspace{-2mm}

\begin{itemize}
    \item \textbf{Title Page:} This section provides a description of the lesson topic, including the key focus areas and the learning objectives the lesson aims to achieve. 
    \item \textbf{Scenario I Section:} In this section, tutors are presented with the first tutoring scenario and are required to answer open-ended questions and multiple-choice questions (MCQ). These questions are designed to replicate real-world tutoring situations, asking tutors to respond as they would during an actual session and to justify their responses. 
    \item \textbf{Instruction:} This section offers research-backed instruction for effective tutoring practices, helping tutors understand both desired and undesired tutoring practice. Each instruction section includes specific examples that help clarify best practices. 
    \item \textbf{Scenario II Section:} This section follows a structure similar to the \textbf{Scenario I} section but presents a different scenario that requires tutors to apply their learned skills in a slightly altered context. This variation ensures that tutors understand the applicability of concepts beyond a single situation. 
    \item \textbf{Conclusion:} The lesson concludes with a summary that reviews the key concepts covered throughout the lesson, reinforcing the understanding of tutoring strategies discussed. The conclusion section also includes a list of references, providing tutors with additional resources for deeper exploration
    \vspace{-3mm}
\end{itemize}

\subsection{Human-Crafted Lesson}
\vspace{-1mm}
We invited two lesson designers to craft the lessons for tutor training and both lesson designers followed the structured lesson format outlined in Section \ref{lesson_structure}. Each designer was tasked with creating lessons aligned with specific tutoring scenarios. The first lesson designer developed two lessons. The first lesson is \textit{``Encouraging Students' Independence''} which focuses on fostering student autonomy by training tutors to use polite, constructive language to guide learners without undermining their independence and the second lesson is \textit{``Turning on Cameras''}, which aims to equip novice tutors with strategies to encourage camera use in online tutoring sessions, aiming to improve student engagement and active participation. The second lesson designer developed one lesson, i.e., \textit{``Encouraging Help-Seeking Behavior''}, which aims to train tutors in identifying and addressing barriers that prevent students from seeking help, fostering a more supportive learning environment. All three lessons underwent rigorous review and multiple iterations, ensuring they were polished and ready for publication on the tutor training platform. The topics of these three human-crafted lessons were subsequently used to prompt the LLMs, following the prompt engineering approaches detailed in Section \ref{prompt_engineering}. The human-crafted lessons are publicly accessible through the  \href{https://osf.io/wafnm/files/osfstorage?view_only=e6dfcb2c31854245b4882621f7469be5}{OSF repository}
\vspace{-3mm}

\subsection{Large Language Model-Generated Lessons}
\label{prompt_engineering}
\begin{figure}[btp]
\centering
\includegraphics[width=12cm]{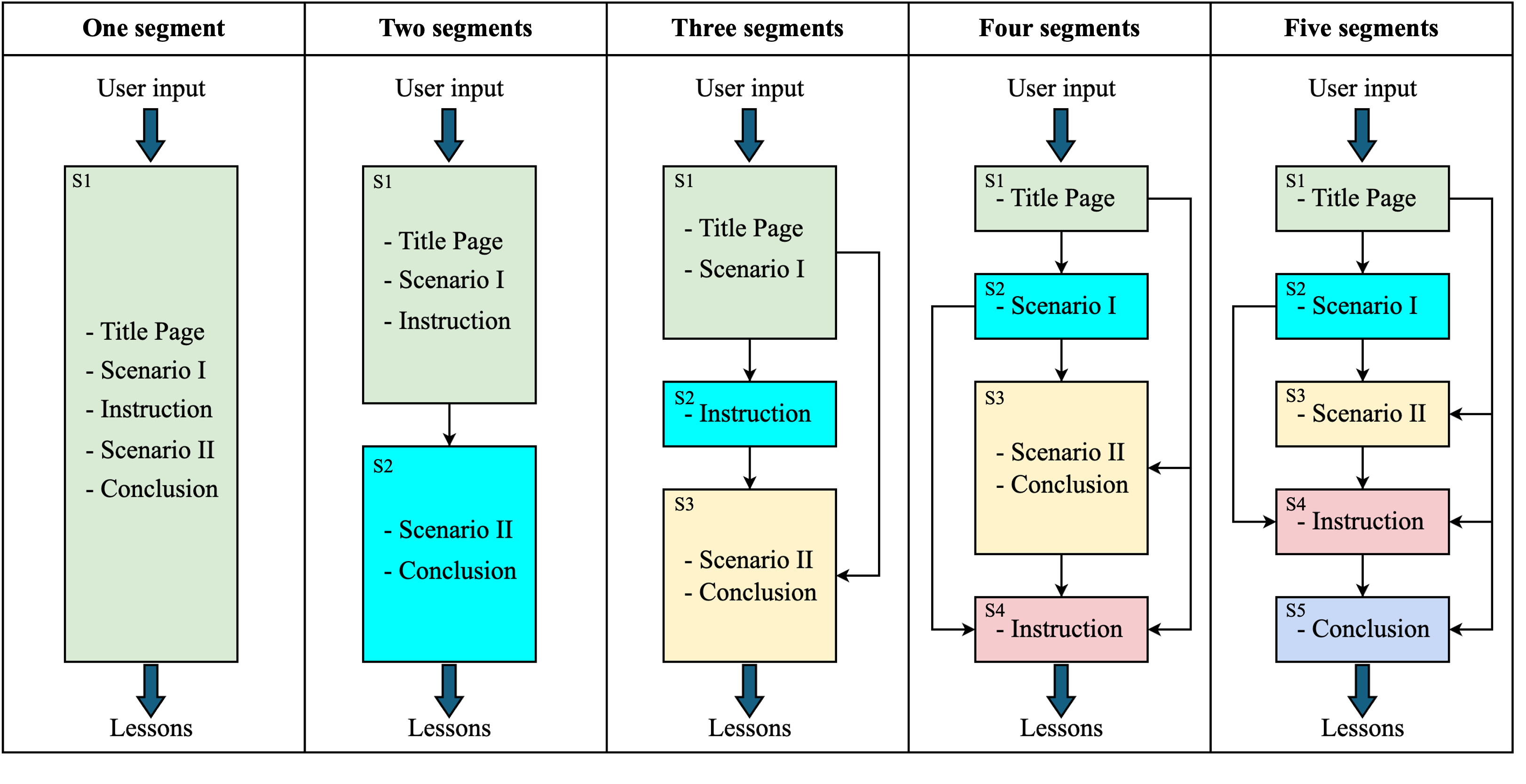}
\vspace{-3mm}
\caption{Our approaches for prompting LLM to generate lessons for tutor training.}
\label{lesson_generation}
\vspace{-3mm}
\end{figure}

We aim to generate comprehensive tutor training lessons based on research articles related to effective tutoring practices. For generating more topic-specific content, Retrieval-Augmented Generation (RAG) has emerged as a powerful technique \cite{lewis2020retrieval, gao2023retrieval}. RAG-based LLM content generation combines both retrieval and generation capabilities, allowing the model to first retrieve relevant documents and then utilize the relevant information to generate content. The strength of RAG lies in its ability to produce contextually accurate content by incorporating the generated output into factual data, which is particularly advantageous when generating educational content that requires specific knowledge from reliable sources \cite{rao2024ramo}. As shown in Figure \ref{lesson_generation}, we propose a two-step approach for lesson generation. Firstly, a lesson designer retrieves articles related to specific tutoring practices, providing the foundational knowledge on effective instructional strategies. Secondly, these retrieved articles are used to prompt the GPT-4o model (\texttt{gpt-4o-2024-05-13}). The GPT-4o model has been recognized for its balance of cost efficiency and strong performance in content generation tasks, making it a suitable choice for producing pedagogically sound lesson content.

An important aspect of generating tutor training lessons is determining the optimal prompting strategy for content generation. As discussed in Section \ref{lesson_structure}, each lesson consists of five main sections. 
  It is important to explore whether it is more effective to generate an entire lesson in one step or generate multiple segments of lesson section (e.g., generate each section independently). 
  As shown in Figure \ref{lesson_generation}, the lesson structure includes five sections, and each section can be generated using different prompting strategies. Additional details about our prompting strategies can be accessed through the  \href{https://osf.io/wafnm/files/osfstorage?view_only=e6dfcb2c31854245b4882621f7469be5}{OSF repository}



\begin{itemize}
    \item \textbf{One Segment:} The entire lesson (all five sections) is generated at once. 
    \item \textbf{Two Segments:} The lesson is split into two main segments for generation. 
    The content from S1, including the title page and the description of Scenario I, is incorporated as part of the prompt for generating S2. 
    \item \textbf{Three Segments:} The lesson was divided into three segments. 
    The content from S1 is used to prompt the generation of the instructional content (S2), while both S1 and S2 are incorporated as inputs to generate Scenario II and the conclusion (S3). 
    \item \textbf{Four Segments:} The lesson was split into four segments. 
    Since the instruction section includes descriptions of research-recommended tutoring practices, generating it independently enhances clarity and depth. Scenario II and the Conclusion are generated together to streamline transitions. 
    \item \textbf{Five Segments:} The most granular approach involves generating each of the five segments. 
    This approach provided the highest level of detail and customization for each section, allowing the model to focus exclusively on the objectives of a single segment. 
\end{itemize}


\subsection{Lesson Evaluation}

To evaluate the influence of different prompting strategies for lesson generation (RQ1), we hired two human coders to rate the content generated by the LLM. Both coders were experienced in tutoring middle school math subjects and had completed over ten tutor training lessons on the tutor training platform. 
The coders followed a detailed rating rubric, which was summarized from previous research \cite{vai2011essentials} and included 17 distinct codes for evaluating lesson quality. Due to the page limit, the rating rubric is illustrated in Table A1 in \href{https://osf.io/7wfg9?view_only=e6dfcb2c31854245b4882621f7469be5}{Digital Appendix}. To ensure rating consistency, both coders participated in multiple training sessions where they collaboratively rated a sample of LLM-generated lessons. These sessions continued until the coders achieved sufficient agreement on the application of the codes. After training, the two coders independently rated the LLM-generated lessons. The inter-rater reliability was assessed using Cohen's $\kappa$, which resulted in a score of 0.72. This score indicates substantial agreement \cite{neuendorf2017content}.  Any discrepancies were addressed through discussion involving both coders and a third reviewer, who also had experience tutoring middle school math and had completed over ten tutor training lessons on the platform. This process ensured that the final ratings accurately reflected the quality of the generated lessons. 
To answer RQ2, we collected feedback from two lesson designers using a structured written questionnaire. The designers were provided with a set of open-ended questions (see Table A2 in \href{https://osf.io/7wfg9?view_only=e6dfcb2c31854245b4882621f7469be5}{Digital Appendix}). Each designer independently provided written responses to all questions.

\vspace{-3mm}



\section{Results}
\subsection{Analyze the quality of LLM-generated tutor training lessons}

The results of the human rating scores (Table \ref{table:rq1}) demonstrate that the three-segment approach (\textbf{Three Seg.}) achieved the highest average rating (14.67) for lesson generation, while the one-segment approach (\textbf{One Seg.}) received the lowest average rating (10.67). These findings indicate that decomposing the lesson generation task into multiple subtasks can enhance the quality of the generated content. 
Interestingly, the five-segment approach (\textbf{Five Seg.}), which represents the highest level of task decomposition, performed slightly worse than the three-segment approach.

\vspace{-4mm}

\begin{table}[ht]
\centering
\scriptsize
\renewcommand{\arraystretch}{1.35}
\caption{Rating scores for LLM-generated tutor training lessons in different segments.}
\label{table:rq1}
\begin{tabular}{lccccc}
\hline
\textbf{Lessons}                   & \textbf{One Seg.} & \textbf{Two Seg.} & \textbf{Three Seg.} & \textbf{Four Seg.} & \textbf{Five Seg.} \\ \hline
Encouraging Students' Independence & 12                   & 11                    & 13                      & 13                     & 13                     \\
Encouraging help-seeking behavior  & 8                    & 11                    & 15                      & 15                     & 12                     \\
Turning on Cameras                 & 12                   & 14                    & 16                      & 14                     & 15                     \\ \hline
\textbf{Average}                            & 10.67                & 12                    & 14.67                   & 14                     & 13.33                  \\ \hline
\end{tabular}
\vspace{-2mm}
\flushleft\hspace{1em}\scriptsize\textit{Note:} \textbf{Seg.} denotes \textbf{Segment}.
\vspace{-4mm}
\end{table}


 To gain a deeper understanding, we further analyzed the details of the rating scores for each section across the segment approaches. The scores in Table~\ref{table:rq1_details} reflect the final consensus ratings across the three generated lessons for each segmentation approach. For each criterion (e.g., \textit{Specific} or \textit{Clarity}), a score of 3/3 indicates that all three lessons met the standard for that code, while a score of 0/3 means that none did. In the \textbf{Title page} (see Table \ref{table:rq1_details}), across all segments, the code \textit{Specific} achieved consistently high scores, indicating that the generated content effectively specified what the trainees would be able to do and under what conditions they would demonstrate the behavior. However, significant variability was observed in the code \textit{Clarity}. In particular, the language used to define learning objectives often lacked clarity in certain segment configurations, such as the one-segment, two-segment, and five-segment approaches.

\begin{table}[t]
\centering

\caption{Consensus ratings for each lesson section across different segmentation}
\label{table:rq1_details}
\scriptsize
\renewcommand{\arraystretch}{1.3}
\begin{tabular}{p{3.6cm}>{\centering\arraybackslash}p{1.5cm}>{\centering\arraybackslash}p{1.5cm}>{\centering\arraybackslash}p{1.5cm}>{\centering\arraybackslash}p{1.5cm}>{\centering\arraybackslash}p{1.5cm}}
\hline
\textbf{Evaluation Codes}  & \textbf{One Seg.}    & \textbf{Two Seg.}    & \textbf{Three Seg.}  & \textbf{Four Seg.}   & \textbf{Five Seg.}  \rule{0pt}{3.2ex}\rule[-2.2ex]{0pt}{0pt} \\ \hline
\textbf{Title page}               & \multicolumn{1}{l}{} & \multicolumn{1}{l}{}  & \multicolumn{1}{l}{}    & \multicolumn{1}{l}{}   & \multicolumn{1}{l}{}   \\
\hspace{1em}\textit{Specific}                          & 3/3                  & 3/3                   & 3/3                     & 3/3                    & 3/3                    \\
\hspace{1em}\textit{Clarity}                           & 1/3                  & 0/3                   & 2/3                     & 2/3                    & 1/3                    \\ 
\textbf{Scenario 1}                        &                      &                       &                         &                        &                        \\
\hspace{1em}\textit{Problem Structure}                 & 3/3                  & 3/3                   & 3/3                     & 3/3                    & 3/3                    \\
\hspace{1em}\textit{Alignment with LO}                 & 2/3                  & 3/3                   & 3/3                     & 2/3                    & 2/3                    \\
\hspace{1em}\textit{Feedback}                          & 1/3                  & 1/3                   & 3/3                     & 2/3                    & 2/3                    \\
\hspace{1em}\textit{Plain Language}                    & 3/3                  & 3/3                   & 3/3                     & 3/3                    & 3/3                    \\ 
\textbf{Instruction}              &                      &                       &                         &                        &                        \\
\hspace{1em}\textit{Alignment with LO}                 & 3/3                  & 3/3                   & 3/3                     & 3/3                    & 3/3                    \\
\hspace{1em}\textit{Clarity of Writing}                & 0/3                  & 0/3                   & 1/3                     & 2/3                    & 2/3                    \\
\hspace{1em}\textit{Plain Language}                    & 1/3                  & 2/3                   & 3/3                     & 3/3                    & 3/3                    \\
\hspace{1em}\textit{Use of Examples}                   & 3/3                  & 3/3                   & 3/3                     & 3/3                    & 3/3                    \\
\hspace{1em}\textit{Pedagogy Grounded}                 & 0/3                  & 1/3                   & 3/3                     & 1/3                    & 1/3                    \\ 
\textbf{Scenario 2}               &                      &                       &                         &                        &                        \\
\hspace{1em}\textit{Problem Structure}                 & 3/3                  & 3/3                   & 3/3                     & 3/3                    & 3/3                    \\
\hspace{1em}\textit{Alignment with LO}                 & 2/3                  & 3/3                   & 3/3                     & 3/3                    & 3/3                    \\
\hspace{1em}\textit{Feedback}                          & 1/3                  & 2/3                   & 2/3                     & 2/3                    & 2/3                    \\
\hspace{1em}\textit{Plain Language}                    & 3/3                  & 3/3                   & 3/3                     & 3/3                    & 3/3                    \\ 
\textbf{Conclusion}               &                      &                       &                         &                        &                        \\
\hspace{1em}\textit{Alignment with LO}                 & 3/3                  & 3/3                   & 3/3                     & 3/3                    & 3/3                    \\
\hspace{1em}\textit{Authenticity of References}        & 0/3                  & 0/3                   & 0/3                     & 1/3                    & 0/3                    \\ \hline
\end{tabular}
\vspace{-2mm}
\flushleft\hspace{1em}\scriptsize\textit{Note:} \textbf{Seg.} denotes \textbf{Segment}. Each score (e.g., 3/3) represents the number of generated lessons (out of three) that received a positive consensus rating for the given criterion and segmentation. 
\vspace{-4mm}
\end{table}



For \textbf{Scenario 1} and \textbf{Scenario 2}, the code \textit{Problem Structure} consistently received high ratings across all segment prompts, highlighting the model's ability to replicate the structured format of tutor training lessons effectively. Additionally, both scenarios demonstrated strong alignment with learning objectives and employed clear, accessible language, as reflected in the high ratings for the codes \textit{Alignment with Learning Objectives (LO)} and the \textit{Plain Language} across all segment prompts. The provision of feedback (code \textit{Feedback}) showed notable improvements in prompts using three or more segments compared to the one- or two-segment approaches. This improvement is likely attributed to the decomposition of content generation tasks, which enables the LLM to allocate more attention to generate feedback-related content


In the \textbf{Instruction} section, the LLM-generated lessons also demonstrated strong alignment with the learning objectives and provided practical examples. However, the code \textit{Clarity of Writing} received lower scores for the one- and two-segment approaches. This limitation likely stems from the generation process, where the instruction section was integrated into a larger task that combined multiple components (see Figure \ref{lesson_generation}). In contrast, the instruction section was generated independently in the three-, four-, and five-segment approaches, which allowed the LLM to focus more on delivering specialized concepts with detailed explanation derived from research-based articles. The code \textit{Pedagogy Grounded} also showed variability across segment configurations, receiving lower scores across all approaches except the three-segment setup. Pedagogical grounding, which assesses the integration of theoretical principles into practical instructional guidance, was most effective in the three-segment approach. This improvement might be attributed to the sequential generation process, where the instruction section was generated earlier in the three-segment strategy compared to the four- and five-segment approaches. Further investigation is required to better understand the underlying reasons for the lower scores in \textit{Clarity of Writing} and \textit{Pedagogy Grounded}, which will help refine the prompting strategies to achieve more consistent performance across all sections of the lesson.


Finally, the LLM-generated lesson section \textbf{Conclusion} consistently underperformed in generating authentic references across all segment configurations, as reflected by the low scores for the code \textit{Authenticity of References} in Table \ref{table:rq1_details}. Even with the optimal segment configuration (i.e., three segments) combined with the RAG approach—which integrated academic papers containing verified references—the model still produced references that did not exist (see Table A3 in \href{https://osf.io/7wfg9?view_only=e6dfcb2c31854245b4882621f7469be5}{Digital Appendix}). This finding highlights that while the RAG approach enhances the contextual relevance of generated content by incorporating retrieved documents, it does not inherently address the issue of citation authenticity. 

\vspace{-3mm}

\subsection{Compare LLM-generated Lessons with Human-Crafted Lessons}

To answer \textbf{RQ2}, we invited two lesson designers who were asked to identify strengths and weaknesses of these generated lessons compared to their manually crafted lessons, which encouraged critical analysis rather than subjective preference. 
Building on the findings from RQ1, we selected lessons generated using the three-segment prompt for comparison, as it generally demonstrated optimal performance across different lessons and codes as shown in Table \ref{table:rq1} and \ref{table:rq1_details}.  The reflections of lesson designers revealed several positive aspects of LLM-generated lessons, highlighting the system's utility in lesson design.

\vspace{-3mm}
\subsubsection{Strengths of LLM-generated Lesson}
\label{strengths_llm}
\begin{figure}[btp]
\centering
\includegraphics[width=10.5cm]{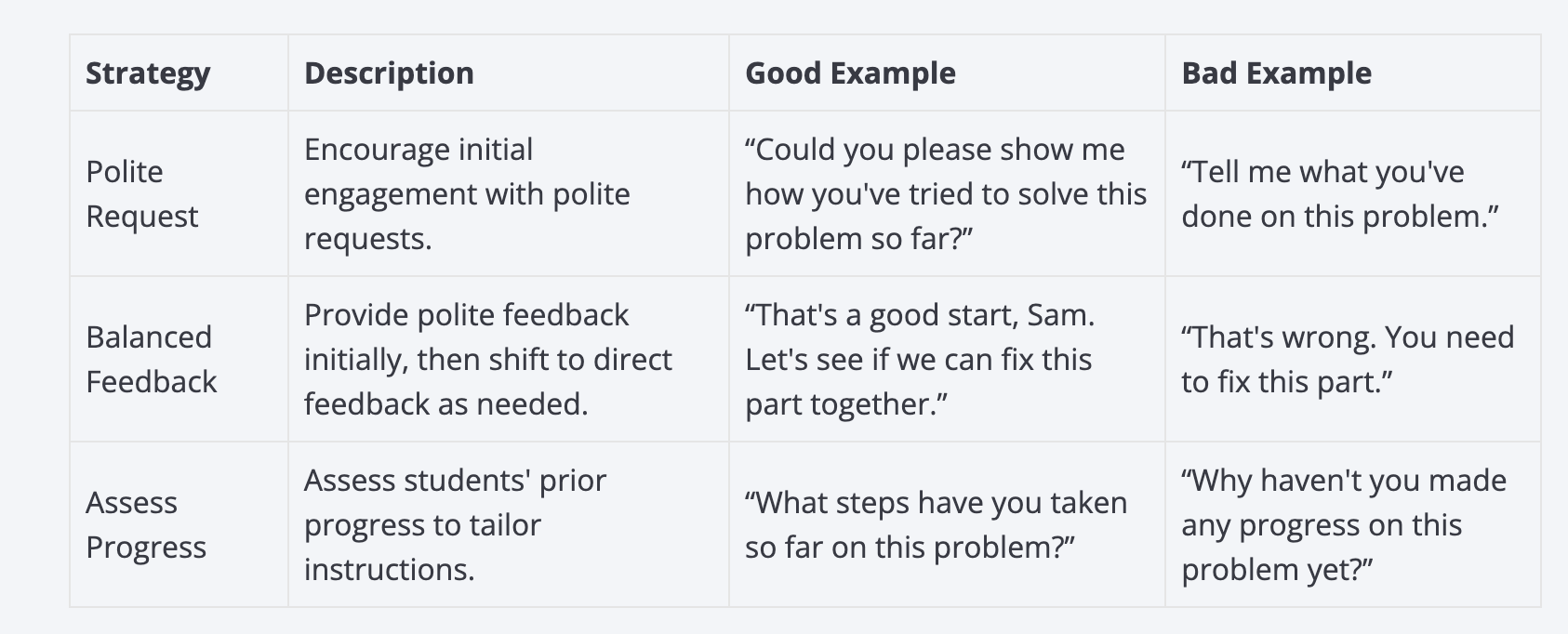}
\vspace{-2mm}
\caption{LLM-generated examples of  strategies for the lesson \textit{Encouraging Students' Independence}. The chart highlights three strategies—Polite Request, Balanced Feedback, and Assess Progress—along with their descriptions, good examples of implementation, and corresponding bad examples to avoid.}
\label{the_chart}
\vspace{-5mm}
\end{figure}

\textbf{Time-saving in creating lesson:} One of the most significant advantages of LLM-facilitated lesson generation, as highlighted by both lesson designers, is the substantial time savings in drafting scenario sections. Crafting effective scenarios is a difficult process that requires maintaining a careful balance between similarity and distinctiveness across questions. As described in Section \ref{lesson_structure}, Scenario I and Scenario II in a lesson must be comparable in difficulty and context to provide consistent learning opportunities, yet distinct enough to prevent students from perceiving them as identical. 
Lesson designer 1 elaborated on the challenge, noting, \textit{``...The process of authoring and adjusting is incredibly time-consuming. The AI can quickly generate variations of scenarios, allowing educators to focus more on fine-tuning and less on repetitive drafting...''} Similarly, Lesson designer 2 emphasized the benefit of LLM assistance in creating multiple-choice questions within these scenarios, stating, \textit{``...I think the LLM-generated lessons do save a lot of time from researching and creating the multiple choice questions.''} Both designers agreed that this time-saving feature significantly reduces the cognitive and logistical load associated with scenario design. Additionally, designer 1 further commend \textit{``...the instructional charts that's generated are mostly very helpful. ...''} as shown in Figure \ref{the_chart}. The feedback from the two lesson designers suggests that the LLM-generated lessons can enhance the lesson creation workflow.

\textbf{Generating diverse and real-world scenarios:} Lesson designer 1 noted, \textit{``The AI can quickly generate variations of scenarios,''} and \textit{``I think the scenarios are generally very-well generated. When given a strategy, GPT seems to be able to give scenarios where that strategy can be used, which I think is pretty helpful''} highlighting the model's capability to provide multiple scenarios tailored to different instructional strategies. For example, in the lesson focused on Encouraging Students' Independence, the LLM-generated content in both scenario sections: 
\vspace{-1.5mm}
\begin{itemize}
    \item \textbf{Scenario I:} \textit{``Tutor Alex is working with a student named Jamie, who often waits for direct answers rather than attempting to solve math problems independently. Jamie has shown some progress but still lacks confidence in tackling problems alone.''} 
    \item \textbf{Scenario II:} \textit{``Tutor Taylor is working with a student named Sam, who frequently asks for hints and confirmations before making any progress on math problems. Sam has shown the ability to solve problems but often seeks reassurance, which hinders independent problem-solving.''} 
    \vspace{-1.5mm}
\end{itemize}



The examples above demonstrate how subtle variations in the scenarios can address distinct aspects of encouraging independence—one focusing on building confidence in hesitant students (\textbf{Scenario I}) and the other on managing students' reliance on reassurance (\textbf{Scenario II}). 

\textbf{Unidentified biased or offensive content:} 
Ensuring unbiased and appropriate AI-facilitated learning content generation is essential for building trust among educators and learners. Lesson designer 1 highlighted this positive aspect, stating, \textit{``I haven’t seen any biased or offensive content in GPT-generated lessons.''} Similarly, Lesson Designer 2 affirmed, \textit{``I haven’t seen this,... do not seem to include biased or offensive material.''} The feedback from both lesson designers suggests that the LLM-generated lesson has avoided inappropriate content. 

\textbf{Facilitating Human-AI Collaboration:} As acknowledged by both lesson designers, the AI-generated content provided a foundational draft that could be iteratively improved, significantly reducing the time and effort required to create lessons from scratch. Lesson designer 1 emphasized the value of this collaboration, noting, \textit{``...The AI can quickly generate variations of scenarios, allowing educators to focus more on fine-tuning and less on repetitive drafting...''} and similarly, lesson designer 2 indicated, \textit{``The quality of some LLMs generated lessons are good since they can be used after a few rounds of modification. The lessons I create also have been reviewing and polishing for several rounds...''} These observations underscore the collaborative potential of LLMs in lesson preparation and highlight a synergistic approach: AI efficiently handles repetitive and time-intensive tasks, while human expertise ensures the final output aligns with instructional best practices and meets the unique needs of learners. 

\vspace{-3mm}

\subsubsection{Weaknesses of LLM-Generated Lessons}


While the positive aspects of LLM-facilitated lesson generation highlight its potential to streamline the process, reduce repetitive tasks, and generate realistic scenarios, it also has limitations. Lesson designers identified several areas where AI-generated content falls short of the standards achieved in human-crafted lessons.

\textbf{Generic feedback content:} Unlike handcrafted lessons, which often include detailed and targeted feedback for each possible answer, the AI-generated feedback content mostly focused on explaining the correct answer. This approach overlooked the importance of addressing why specific options were incorrect, leaving gaps in the learning experience. Lesson designer 1 noted, \textit{``...Handcrafted questions often provide targeted feedback for each possible answer, explaining why a specific choice is correct or incorrect. In contrast, LLM-generated courses tend to focus only on explaining the correct answer, neglecting the nuanced reasoning behind other options. In the cases where an explanation for a wrong answer is provided, the explanation given is generally more one-size-fits-all style, which lacks that important factor of targeted feedback...''} Lesson designer 2 indicated \textit{``The explanation for options 2, 3, 4 is not specific enough, since it doesn't explain exactly why we don’t choose them.''} and provided a specific example of LLM-generated feedback (see Table A4 in \href{https://osf.io/7wfg9?view_only=e6dfcb2c31854245b4882621f7469be5}{Digital Appendix}). These observations highlight the necessity of incorporating more nuanced explanations for all answer options, both correct and incorrect. Lacking detailed feedback for incorrect answers diminishes the educational value of the lesson. Targeted feedback is critical for helping learners understand the rationale behind their choices and correct misconceptions. The generic explanations provided by the LLM fail to address this need, resulting in less effective learning experiences.
 


\textbf{Clarity in the title page and instruction section:} Both designers noted instances where the generated lessons inconsistently used terms like ``learners'', ``teachers'', ``tutors'', and ``students''. This inconsistency sometimes confused the intended novice tutors, as the lessons were specifically designed for training tutors on how to teach their students effectively. As shown in Table \ref{tab:lo_examples}, the examples generated by one-segment prompt fail to maintain clarity about the intended novice tutors—confusing ``learners'' with ``tutors''—and does not clearly address the objectives of the tutor training lesson. 
Such inconsistencies can hinder novice tutors' understanding of how the training content applies to their practice.



Additionally, the instruction section of the LLM-generated lessons exhibited similar issues of lack of clarity.  Lesson designer 1 highlighted that \textit{``...the generated instructions are overly long and lack a logical backbone to support the reasoning behind the strategies...''} This disorganization often made it difficult for users to follow the intended flow, as instructions were included excessive details and citations. Additionally, lesson designer 1 added \textit{``...There are some cases where the length and citations used are too much, making it difficult to pick out the emphasis of the instruction...''} An example of this overuse of citations can be seen in Table A3 in \href{https://osf.io/7wfg9?view_only=e6dfcb2c31854245b4882621f7469be5}{Digital Appendix}. Lesson designer 2 echoed similar concerns, emphasizing the structural and logical disconnect in the instruction sections. \textit{``...The flow and logic can be more natural....''}. This lack of cohesive structure often led to instructions that felt fragmented and harder to comprehend. \textit{``...I find it difficult to read and understand since the logic can be a bit disconnected...''}. An example of this issue can be found in the following excerpt from a generated lesson: ``\textit{...Wallace, Hand, and Prain (2004) highlighted the effectiveness of the Science Writing Heuristic (SWH) in improving conceptual understanding through student-generated questions...}'' The lack of context or explanation about what the SWH framework entails makes it difficult for readers unfamiliar with the framework to understand its relevance. This omission can leave readers, particularly novice tutors, struggling to understand how such frameworks apply to their tutoring practices. 


\begin{table}[t]
\centering
\scriptsize
\caption{Examples of learning objectives generated using three-segment and one-segment prompts}
\label{tab:lo_examples}
\renewcommand{\arraystretch}{1.45}
\begin{tabular}{p{6.5cm}p{5cm}}
\hline
\textbf{Three-segment generation}                                                                                                                                                                                                                                                        & \textbf{One-segment generation}                                                                                                                                                              \\ \hline
\begin{tabular}[c]{@{}p{6cm}@{}}\textit{``Explain how polite instructional strategies can improve student engagement and problem-solving in online math tutoring.''}\\ \textit{``Develop and implement a balanced approach to using polite and direct instructions to foster student independence.''}\end{tabular} & \begin{tabular}[c]{@{}p{5cm}@{}}\textit{``Describe how learners will interpret or explain the tutoring strategy.''}\\ \textit{``Outline how learners will generate or implement the tutoring strategy.''}\end{tabular} \\ \hline
\end{tabular}
\vspace{-5mm}
\end{table}

\vspace{-3mm}

\section{Discussion and Conclusion}
\vspace{-2mm}
Our study investigated the potential of using large language models (LLMs), specifically with the retrieval-augmented generation (RAG) approach, to automate the generation of tutor training lessons for middle-school math pedagogy. Through structured evaluations, we assessed the quality, strengths, and limitations of LLM-generated lessons. Our findings underscore the importance of task decomposition in lesson generation. 
By decomposing the task into manageable subtasks, the LLM could maintain logical coherence across lesson components, demonstrating consistency with prior research advocating task decomposition strategies \cite{wu2022ai, khot2022decomposed}. Conversely, the one-segment approach, which involved generating the entire lesson within a single prompt, resulted in lower-quality outputs. Interestingly, the five-segment approach, representing the highest level of decomposition, introduced challenges such as reduced clarity and weaker pedagogical grounding. These challenges highlight the need for further investigation to determine whether they stem from excessive decomposition, potentially disrupting the logical flow and consistency of the generated content, or from rating bias or manual rating inconsistencies that may have influenced the findings. 
Moreover, our study offers practical insights for using LLMs in educational content creation, demonstrating their ability to generate diverse and realistic lesson scenarios, establish the ethical reliability of the generated lessons, and support their value in scaling inclusive educational resources.

Despite these advantages, our investigation identified key areas where LLM-generated content requires further improvement. Feedback provided for open-ended and multiple-choice questions frequently lacked detailed explanations necessary for addressing misconceptions related to incorrect responses. Furthermore, issues with clarity were evident in the lesson content including title page (e.g., unclear descriptions of learning objectives) and instructional sections (e.g., logical disconnects, and insufficient explanation of jargon). These shortcomings highlight the need for refined prompting strategies to improve precision and coherence in generated content. Another weakness of LLM-generated lessons is the authenticity of academic references, which were often unreliable. Human lesson designers still need to manually verify and supplement references if lessons require source citations. The strengths and weaknesses of LLM-generated lessons emphasize that while LLMs hold immense potential for automating lesson generation, their effectiveness is maximized within a human-AI collaborative workflow. Educators and instructional designers play a crucial role in reviewing and refining AI-generated content to ensure it aligns with pedagogical objectives and meets the nuanced needs of learners.

\noindent\textbf{Future Works.} Future research should focus on designing more targeted and specific technical approaches to address limitations identified in LLM-generated lessons. For instance, a multi-agent approach could assign specialized agents to generate, evaluate, and refine lesson content, thereby improving feedback quality through iterative review \cite{cao2025first}. Moreover, integrating multiple modalities—such as pairing AI-generated text with relevant visuals or slides—into lessons could better support diverse learning preferences in tutor training \cite{zhao2025slideitright}. Additionally, our study's lesson evaluation rubric was primarily informed by one resource  \cite{vai2011essentials}. While the rubric \cite{vai2011essentials} is comprehensive, future research could benefit from comparing or integrating it with established rubrics in the field to further validate and strengthen lesson evaluation. Then, enabling direct collaboration between instructional designers and the LLM during the refinement process can provide opportunities for real-time adjustments. For example, instructional designers could request immediate changes to tone, clarity, or content specificity, iteratively improving lesson quality. This interactive feedback loop would increase the quality of generated content and streamline the lesson creation process. Lastly, our study primarily focused on evaluating LLM-generated lessons based on human evaluators' judgments, serving as a foundational step in identifying high-quality lessons for further exploration. Future research should conduct experiments to evaluate the relative effectiveness of LLM-generated lessons compared to human-crafted ones, providing insights into their impact on participants' learning outcomes. The results would help clarify the extent to which LLM-generated lessons achieve comparable or complementary results to those designed by human experts.

\section*{Acknowledgment} This work was supported by the Learning Engineering Virtual Institute,\footnote{ \href{https://learning-engineering-virtual-institute.org/}{https://learning-engineering-virtual-institute.org/}} and a grant from the University Research Committee (Grant No. 2401102970) at The University of Hong Kong. The opinions, findings, and conclusions expressed in this paper are solely those of the authors.

\vspace{-4mm}
\bibliographystyle{splncs04}
\bibliography{mybibliography}

\end{document}